\documentclass[11pt]{article}
\usepackage{amssymb,epsfig,cite}

\renewcommand{\baselinestretch}{1.62}

\makeatletter
\def\@cite#1#2{{\footnotesize$^{\bf #1}$\if@tempswa , #2\fi}}
\renewcommand{\@biblabel}[1]{#1.}

\long\def\@makecaption#1#2{%
  \vskip\abovecaptionskip
  \sbox\@tempboxa{#2}%
  \ifdim \wd\@tempboxa >\hsize
    #2\par
  \else
    \global \@minipagefalse
    \hb@xt@\hsize{\hfil\box\@tempboxa\hfil}%
  \fi
  \vskip\belowcaptionskip}
\makeatother

\begin{document}
\noindent
\LARGE
\textbf{Magnetic trapping of neutrons}
\normalsize
\vspace{\baselineskip}

\renewcommand{\baselinestretch}{1.1}
\noindent
\textbf{P.\,R.~Huffman$^{*\dag}$,
        C.\,R.~Brome$^{*}$,
        J.\,S.~Butterworth$^{*\ddag}$,
        K.\,J.~Coakley$^{\S}$, \\
	M.\,S.~Dewey$^{\dag}$,
	S.\,N.~Dzhosyuk$^{*}$,
        R.~Golub$^{\parallel}$,
        G.\,L.~Greene$^{\P}$,
        K.~Habicht$^{\parallel}$, \\
        S.\,K.~Lamoreaux$^{\P}$,
        C.\,E.\,H.~Mattoni$^{*}$,
        D.\,N.~McKinsey$^{*}$,
	F.\,E.~Wietfeldt$^{*\dag}$, \\
        \& J.\,M.~Doyle$^{*}$
        }
\addtocounter{footnote}{2}
\renewcommand{\thefootnote}{\fnsymbol{footnote}}
\footnotetext[3]{Present address: Institut Laue-Langevin, B.P. 156-6 rue Jules 
	Horowitz, 38042, Grenoble (Cedex 9), France.}
    
\footnotesize
\mbox{~} \\
\noindent
    ${*}$~Harvard University, 17 Oxford Street, Cambridge, Massachusetts 02138, USA.\\
    ${\dag}$~National Institute of Standards and Technology, 100
    	Bureau Drive, MS 8461, Gaithersburg, Maryland 20899, USA.\\
    ${\S}$~National Institute of Standards and Technology, 325
    	Broadway, MS 898.02, Boulder, Colorado 80303, USA.\\
    ${\parallel}$~Hahn-Meitner-Institut, Glienicker Stra\ss\/e 100,
    	D-14109 Berlin, Germany.\\
    ${\P}$~University of California, Los Alamos National Laboratory,
    	P.O. Box 1663, Los Alamos, New Mexico 87545, USA. \\
    \mbox{~}
	
\renewcommand{\baselinestretch}{1.62}
\normalsize
\hrule
\mbox{~} \\
\noindent\textbf{Accurate measurement of the lifetime of the neutron
(which is unstable to beta decay) is important for understanding the
weak nuclear force\cite{Dubbers} and the creation of matter during the
Big Bang\cite{Turner}.  Previous measurements of the neutron lifetime
have mainly been limited by certain systematic errors; however, these
could in principle be avoided by performing measurements on neutrons
stored in a magnetic trap\cite{Doyle94}.  Neutral and charged particle
traps are widely used tool for studying both composite and elementary
particles, because they allow long interaction times and isolation
from perturbing environments\cite{opticaltrap}.  Here we report the
magnetic trapping of neutrons.  The trapping region is filled with
superfluid $^{\bf 4}$He, which is used to load neutrons into the trap and as
a scintillator to detect their decay.  Neutrons have a lifetime in the
trap of 750$\mathbf{^{+330}_{-200}}$ seconds, mainly limited by their
beta decay rather than trap losses.  Our experiment verifies
theoretical predictions regarding the loading process and
magnetic trapping of neutrons.  Further refinement of this method
should lead to improved precision in the neutron lifetime measurement.}

\makeatletter
\def\@cite#1#2{{\footnotesize$^{#1}$\if@tempswa , #2\fi}}
\makeatother

A more precisely known value of the neutron lifetime will benefit both
weak interaction and Big Bang Nucleosynthesis (BBN) theory.  Combined
with the measured neutron beta-decay asymmetry, the neutron lifetime
yields the $V_{ud}$ element of the Cabibbo-Kobayashi-Maskawa (CKM)
quark mixing matrix.  Non-unitarity of the CKM matrix would indicate
physics beyond the Standard Model\cite{Dubbers}.  In addition, the
theoretical uncertainty in the predicted $^{4}$He abundance in the BBN
model is dominated by the uncertainty in the neutron
lifetime\cite{Turner}.

Many experiments, using different methods, have measured the neutron
beta-decay lifetime\cite{pdg98,BeamLifetime,Mambo,GravTrap,NESTOR}. 
The most precise measurements use ultracold neutrons (UCN, neutrons
with energies $E/k_{b} \leqslant 1$~mK, where $k_{b}$ is Boltzmann's
constant)\cite{ucnbook} confined in a `bottle' made of materials with
high UCN reflectivity.  The largest uncertainties are due to UCN
interactions with the material walls and fluctuations in the initial
filling densities\cite{Mambo,GravTrap}.  Extrapolation methods are
necessary to extract the neutron beta-decay lifetime.  In another
experiment, neutrons with specific trajectories and in a band of
velocities (typically between 10--14~m~s$^{-1}$) were kept in a
magnetic storage ring similar to those used in high-energy particle
physics\cite{NESTOR}.  Since storage rings only confine neutrons in
two dimensions, coupling between the large longitudinal momentum and
small radial momentum by way of betatron oscillations allows neutrons
to escape.  In contrast, neutrons in magnetic traps are not
susceptible to wall losses or betatron oscillations.

Static magnetic traps are formed by creating a magnetic field minimum
in free space.  The confining potential depth ($D$) of such a trap is
determined by the magnetic moment of the trapped species ($\mu$) and
the difference ($\Delta B$) between the magnitude of the field at the
edge of the trap and at the minimum, $D = \mu \Delta B$.  A particle
in a low-field-seeking state (one with its magnetic moment
anti-parallel with the local magnetic field vector) is pushed towards
the trap minimum.  Low-field-seeking particles with total energy less
than $D$ are energetically forbidden from leaving the trapping region. 
For atoms and molecules with a magnetic moment of one Bohr magneton it
is possible to produce trap depths of $\sim 1$~K\@.  The trap depth
for a neutron in the same trap would be only $\sim 1$~mK, because of
its much smaller magnetic moment.  Despite this difficulty, magnetic
trapping of the neutron was proposed as early as 1961 by
Vladimirski\u{i}\cite{vladimirskii}.  His proposed technique was later
used to confine neutrons using a combination of gravity and
magnets\cite{Abo86}.  A separate effort to trap neutrons using a
similar loading method to our work (but different detection scheme)
was unsuccessful because of the high temperature of the helium during
the loading phase\cite{Nie83}.

Crucial to the utility of traps are the techniques used to load them.  In
order to catch a particle in a static conservative trap, its energy
must be lowered while it is in the potential well.  Atoms and
molecules have been cooled and loaded into magnetic traps by
scattering with either cryogenic
surfaces\cite{Wallcooling,Wallcooling2}, cold gases\cite{Doyle95} or
photons from a laser beam (laser cooling)\cite{Chu92}.  Neutrons,
however, cannot be loaded by such methods because they cannot be
excited optically and interact too weakly with atoms to be effectively
cooled by a gas.  Direct thermalization with a cold solid or liquid is
generally precluded by the high probability for neutron absorption in
the vast majority of materials.

Our trapping of neutrons relies on a loading technique that employs
the ``superthermal process''\cite{Golub77}.  A neutron with kinetic
energy near 11~K (where the free neutron and superfluid helium
dispersion curves cross) that passes through the helium-filled
trapping region can lose nearly all of its energy through the creation
of a single phonon.  Neutrons that scatter to energies less than the
trap depth ($\leqslant 1$~mK) and in the appropriate spin state are
trapped.  Upscattering of UCN in liquid helium by single
phonon absorption is highly suppressed due to the low density of 11~K
phonons in the superfluid (proportional to $e^{-11 \mathrm{K}/T}$). 
The rate of higher-order (multiple-phonon) upscattering processes is
proportional to $T^{7}$(ref.~\citen{golub83}).  The helium is maintained
below a temperature of 250~mK to minimize upscattering of the trapped
neutrons (total calculated upscattering rate per neutron $< 
10^{-6}~$s$^{-1}$).

Snother possible loss mechanism for trapped neutrons is nuclear
absorption due to $^{3}$He contamination of the superfluid $^{4}$He
bath.  Unlike $^4$He which does not absorb thermal neutrons at all,
$^3$He readily absorbs neutrons with a thermal cross section, $\sigma
= 5330 \times 10^{-24}$~cm$^2$.  To minimize these losses, our
superfluid-helium bath was isotopically purified ($^3$He/$^4$He $< 5
\times 10^{-13}$; ref.~\citen{Hen87}), giving a calculated absorption
loss rate per neutron of $5 \times 10^{-6}$~s$^{-1}$.  Thus, the
trapped neutrons travel unimpeded through the helium and reside in the
trap for a time limited primarily by the beta-decay lifetime of the
neutron.

\begin{figure}[ht]
\begin{center}
      \epsfig{file=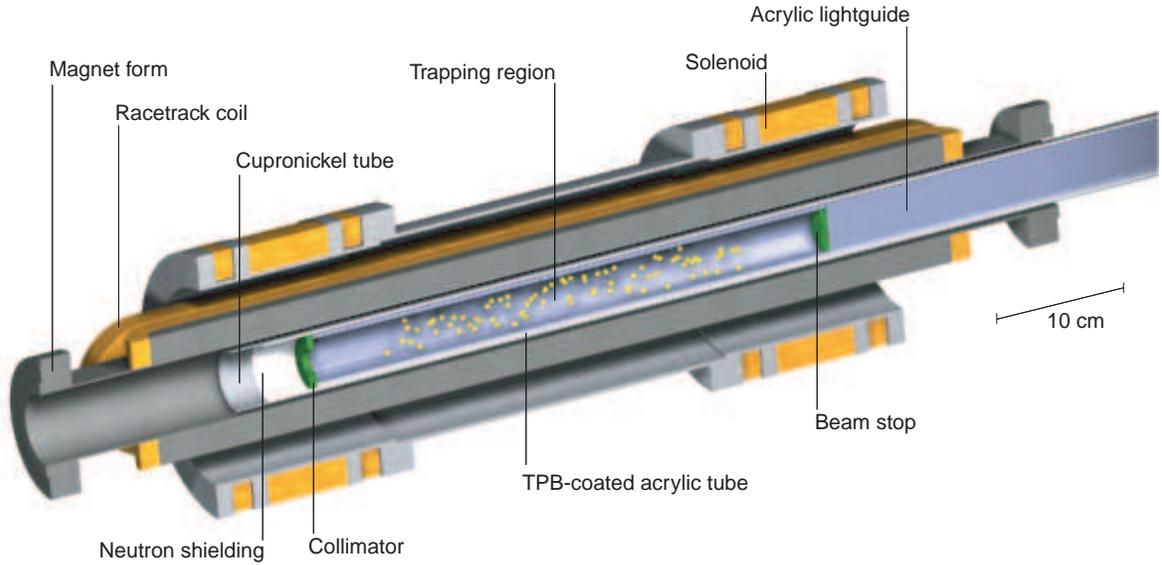}
\end{center}
   \caption{Figure 1:  Half section view of the neutron trapping apparatus.  The
   trapping region is filled with isotopically pure $^{4}$He at a
   temperature $\leqslant 250$~mK\@.  The helium is contained within a
   cupronickel tube.  A beam of cold neutrons passes through a series
   of teflon windows\cite{But98} and enters the helium from the left. 
   It is collimated by a boron carbide ring, passes through the
   trapping region and is absorbed by a boron carbide beam stop. 
   Approximately 1\% of the 11~K neutrons scatter in the superfluid
   helium.  Those neutrons (yellow) in the low-field-seeking spin
   state and with energy below the trap depth are magnetically
   confined.  The rest of the scattered neutrons are absorbed by
   neutron shielding material (boron nitride) surrounding the trapping
   region.  The magnetic trapping field is created by an assembly of
   superconducting magnets.  Radial confinement is provided by a
   quadrupole constructed from four racetrack-shaped coils and axial
   confinement is provided by two sets of solenoids.  Electrons from
   neutron beta-decay cause EUV scintillations in the superfluid
   helium which are wavelength shifted to the visible by a thin film
   of TPB-doped\cite{McK97} polystyrene coated on the inside of an
   acrylic tube surrounding the trapping region.  This tube is
   optically coupled to an acrylic lightguide which transports the
   blue light to the end of the 250~mK region (to the right).}
   \label{fig:apparatus}
\end{figure}

In our experiment, the superfluid helium is contained in a tube that
is axially centered within a superconducting magnetic trap (see
Fig.~\ref{fig:apparatus}).  The incident cold neutron beam is
collimated by a neutron-absorbing ring, passes through the trapping
region, and is absorbed by the beam stop.  As the beam traverses the
trapping region, about 1\% of the 11~K neutrons scatter in the helium. 
Some of these neutrons are trapped and the remainder are absorbed by
shielding materials that surround the helium.

When a trapped neutron decays (into an electron, proton and
anti-neutrino), the resulting high-energy electron (up to 782~keV)
travels through the helium leaving a trail of ionized helium atoms. 
These ions quickly combine with neutral helium atoms to form excimer
diatomic molecules, about half in singlet states and half in triplet
states.  About 15 singlet molecules are created per kiloelectronvolt
of electron energy and the trail is 0--20~mm long\cite{Adams98}.  The
singlet molecules decay within 10~ns, emitting a pulse of light in the
extreme ultraviolet (EUV), $\lambda \approx 70-90$~nm.  This pulse of
scintillation light is the signal of a neutron decaying in the trap. 
Radiative decay of the triplet states has a much longer lifetime,
which we have measured to be $13\pm2$~s (ref.~\citen{McK99}), and does
not contribute to our signal.

The EUV scintillation light is wavelength-shifted to the visible and
transported to the outside of the cryostat where it is detected.  As
shown in Fig.~\ref{fig:apparatus}, the superfluid-helium-filled
trapping region is surrounded by an acrylic tube.  The inside surface
of this tube is coated with a thin layer of polystyrene doped with the
organic fluor tetraphenyl butadiene (TPB; ref.~\citen{McK97}).  The TPB
converts the EUV scintillation light into blue light, a portion of
which is internally reflected down the length of the tube\cite{Hab98}. 
The tube is optically coupled to a solid light guide which transports
the light to the end of the 250~mK region.  The light then passes
through a window at 4~K and into a second light guide which exits the
dewar.  The light is split into two guides that are each coupled to a
photomultiplier tube (PMT).  Background from uncorrelated photons is
suppressed by requiring coincident detection of at least two
photoelectrons in each PMT\@.  This is required because of significant
time-dependent luminescence (initial counting rate without coincidence
detection of $\sim 10^{4}$ s$^{-1}$), induced by neutron absorption in
the shielding materials.  In a series of calibration runs (using a
$^{113}$Sn, 360-keV beta line source placed in the center of the
trapping region) we have determined that a total of $14.5\pm
1.5$~photoelectrons per megaelectronvolt of beta energy are detected
in a pulse 20~ns wide.  We estimate that a coincidence signal is
recorded for $31 \pm 4$\% of neutron decays in the trapping region.

In each experimental run, the cold neutron beam is allowed to pass
through the trapping region for 1,350~s, after which the beam is
blocked and pulses of light are counted for 3,600~s.  While the beam is
on, neutrons interact with the helium and the number of UCN in the trap
increases.  After the beam is turned off, the PMTs are turned on and a
decreasing rate of scintillation events is observed.  A background
signal, consisting of both constant and time-varying components,
obscures the trapped neutron signal.  These backgrounds are subtracted
by taking data in two different kinds of run, trapping (signal +
background) and non-trapping (background only), and subtracting the
two.  In a trapping run the magnetic field is on during the initial
loading phase, so that UCN are confined in the trap, while in a
non-trapping run the magnetic field is initially off, so that no
neutrons can be trapped.  In a non-trapping run, the field is turned
on during detection so that any residual luminescence (which depends
on the magnetic field) will be properly subtracted.  Equal numbers of
each kind of run were taken, pooled and subtracted to give the
background subtracted data.

\begin{figure}[ht] 
\parbox{2.75in}{\epsfig{file=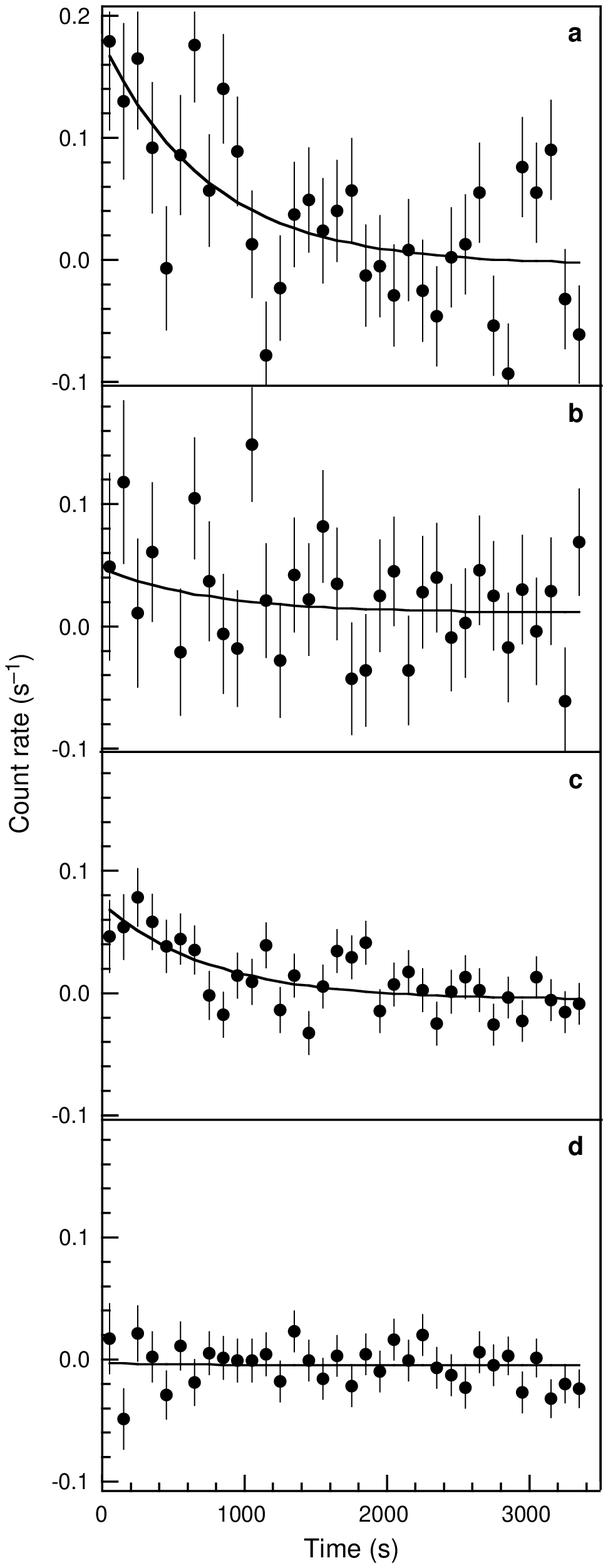,width=2.75 in}}
\parbox{0.125in}{~}
\parbox{3.25in}{\normalsize
    \noindent Figure 2:  Counting rate as a function of time after the neutron
    beam is turned off (pooled background subtracted data).  The two
    sets of trapping data are fit for five parameters: $\alpha_{1}$,
    $\alpha_{2}$, $C_{1}$, $C_{2}$ and $\tau$.  The constant offset
    terms, $C_{1}$ and $C_{2}$ are found to be $-0.004 \pm 0.015$ and
    $-0.006 \pm 0.006$ respectively, both consistent with zero.  The
    best fit values for the initial counting rates for trapping data
    sets 1 and 2 are $\alpha_{1}=0.196 \pm 0.05~\mathrm{s}^{-1}$ and
    $\alpha_{2} =0.084 \pm 0.02~\mathrm{s}^{-1}$.  The detection
    efficiency of $\epsilon = 31 \pm 4$\% combined with the known
    neutron lifetime, gives $N_{1} = 560 \pm 160$ and $N_{2} = 240
    \pm 65$.  The best fit value for the lifetime is $\tau =
    750^{+330}_{-200}$~s.  The results of this fit are shown on plots
    \textbf{a} and \textbf{c}.  The $^{3}$He data is fit to four
    parameters, $\alpha_{1}$, $\alpha_{2}$, $C_{1}$, $C_{2}$, with
    $\tau$ fixed at 750~s.  Because no fit parameter applies to both
    data sets, this is equivalent to two separate two-parameter fits,
    one to each data set.  The amplitudes, $\alpha_{1}^{He}=0.040\pm
    0.045$ and $\alpha_{2}^{He}=0.005\pm 0.016$, are both consistent
    with zero.  The constant offset terms are $C_{1}=0.011\pm 0.011$
    and $C_{2}=-0.006\pm 0.004$.  The $^{3}$He data and fit are shown
    on plots \textbf{b} and \textbf{d}.  (\textbf{a}) Trapping data
    set 1 (23~pairs of runs at a trap depth of 0.76~mK).  (\textbf{b})
    $^{3}$He data set 1 (20~pairs at 0.76~mK).  (\textbf{c}) Trapping
    data set 2 (120~pairs at 0.50~mK).  (\textbf{d}) $^{3}$He data
    set 2 (125~pairs at 0.50~mK).}
    \caption{~}
    \label{fig:data} 
\end{figure}

Two sets of background subtracted data were collected: set 1 with a
trap depth of 0.76~mK (figure~\ref{fig:data}a) and set 2 with a lower
trap depth of 0.50~mK (figure~\ref{fig:data}c).  (The lower trap depth
was used due to problems with the magnet.)  Most of the run-to-run
variation in background rate is eliminated by excluding the first two
pairs of runs in which the background rate is changing quickly due to
activation of materials with lifetimes $> 12$~h.  The remaining
23~pairs in set 1 (from about five days of running) and 120~pairs in
set 2 (from about three weeks of running) are pooled and modeled as:
\begin{equation}
\label{eqn:decayprob} 
W_{1} = \alpha_{1} e^{-t/\tau}+C_{1},~~~~~
W_{2} = \alpha_{2} e^{-t/\tau}+C_{2}. 
\end{equation} 
The subscripts 1 and 2 refer to sets 1 and 2, $\alpha_{i}=\epsilon
N_{i}/\tau$, $N_{i}$ is the initial number of trapped neutrons and 
$i = 1$ or 2,
$\epsilon$ is the detection efficiency, and $\tau$ is the lifetime of
neutrons in the trap.  The constant $C_{i}$ is present to account for
the possible remaining effect of the changing background rate due to
long-lived activation.  However, in all of our fits the value of
$C_{i}$ is consistent with zero.  The fit is performed simultaneously
on the two data sets, minimizing the total $\chi^{2}$ while varying
five parameters: $\alpha_{1}$, $\alpha_{2}$, $C_{1}$, $C_{2}$ and
$\tau$.  The only parameter connecting the two data sets is $\tau$. 
The best fit values indicate $N_{1} = 560 \pm 160$ and $N_{2} = 240
\pm 65$.  Calculations using the known beam flux, trap geometry and
the theory of the superthermal process predict $N_{1} = 480 \pm 100$
and $N_{2} = 255 \pm 50$, in good agreement with the measured values. 
The best fit value for the lifetime, $\tau = 750^{+330}_{-200}$~s is
consistent with the presently accepted value of the neutron beta-decay
lifetime of $886.7 \pm 1.9$~s (ref.~\citen{pdg98}).  All of the errors quoted 
correspond to a 68\% confidence interval.

To verify that our signal is due to trapped neutrons, we doped the
isotopically pure $^{4}$He with $^{3}$He at a concentration of $2
\times 10^{-7}$ $^{3}$He/$^{4}$He.  This amount of $^{3}$He absorbs
the trapped neutrons in less than 1~s without affecting anything else
in the experiment (less than 1\% of the 11~K neutrons are absorbed by
$^{3}$He).  The $^{3}$He background subtracted data (see
Fig.~\ref{fig:data}b and d) is consistent with zero, and differs from
the signal data by more than four standard deviations.  This confirms
that our signal is due to trapped neutrons.

This work demonstrates the loading, trapping, and detection techniques
necessary for performing a neutron lifetime measurement using
magnetically trapped UCN\@.  Another important result is the direct
confirmation of the theoretical prediction of the UCN density in the
trap.  Our value for the number of trapped neutrons at a 0.76~mK trap
depth corresponds to a density of 2~UCN/cm$^{3}$, compared to the
density of 1~UCN/cm$^{3}$ obtained in the UCN material bottle
experiments\cite{Mambo,GravTrap} with a comparable UCN cut-off energy
and higher flux reactor.  Our measured density is consistant with
previous measurements of the UCN production rate based on observation
of upscattered UCN at higher temperatures\cite{kilvington} and 
confirms that the low number of UCN detected in ref.~\citen{Yoshiki} 
was not due to an intrinsically low UCN production rate.

Magnetic trapping should allow significant improvement in the
measurement of the neutron lifetime $\tau_{n}$.  There are several
loss mechanisms specific to this method, including thermal
upscattering and $^{3}$He absorption.  Both of these can be suppressed
to a high level as discussed above.  Another possible loss mechanism
is neutron depolarization.  When the neutron passes through a region
where the field changes direction quickly compared to the neutron's
spin precession frequency, the neutron can flip spin and be ejected
from the trap.  In our trap geometry, the trap has no zero-field
regions and this loss rate can be minimized by the application of a
bias field\cite{Doyle94,Happer}.

Using our method with a higher-flux source of cold neutrons, a
$10^{-5}$ measurement of $\tau_{n}$ should be possible\cite{Doyle94}. 
In the near term, improvements such as increased trap size and depth
(resluting in an improved ratio of signal-to-background events) should
allow an improved accuracy measurement of $\tau_{n}$ at the $5 \times
10^{-4}$ level.  Other potential applications of superthermal UCN
sources include searches for time-reversal violation\cite{EDM} and new
methods of neutron scattering especially suited for the study of large
(biological-scale) molecules\cite{condmatter}.

\vspace{0.5\baselineskip}
\noindent
\footnotesize
Received 5 August; accepted 16 November 1999.
\normalsize
\hrule
\footnotesize \renewcommand{\baselinestretch}{1.1}

\renewcommand{\baselinestretch}{1.62}

\noindent \textbf{Acknowledgments.} We thank J.\,M.~Rowe, 
D.\,M.~Gilliam, G.\,L.~Jones, J.\,S.~Nico, N.~Clarkson, G.\,P.~Lamaze, 
C.~Chin, C.~Davis, D.~Barkin, A.~Black, V.~Dinu, J.~Higbie, H.~Park, 
R.~Ramakrishnan, I.~Siddiqi, and G.~Brandenburg for their help with 
this project.  We thank P.~McClintock, D.~Meredith and P.~Hendry for 
supplying the isotopically pure helium.  We acknowledge the support of 
the NIST, US DOC, in providing the neutron facilities used in this 
work.  This work is supported in part by the US NSF.  The NIST 
authors acknowledge the support of the US DOE.
\vspace{0.5\baselineskip}

\noindent
Correspondence and requests for materials should be addressed to 
P.R.H. (e-mail: paul.huffman@nist.gov).

\pagebreak

\pagebreak

\end{document}